\documentclass[10pt,A4]{article}
\usepackage[top=0.85in,left=1in,footskip=0.75in]{geometry}

\usepackage{amsmath,amssymb}

\usepackage{changepage}

\usepackage[utf8x]{inputenc}

\usepackage{textcomp,marvosym}

\usepackage{natbib}

\usepackage{nameref,hyperref}

\usepackage{microtype}
\DisableLigatures[f]{encoding = *, family = * }

\usepackage[table]{xcolor}

\usepackage{array}

\newcolumntype{+}{!{\vrule width 2pt}}

\newlength\savedwidth

\usepackage{setspace} 
\raggedright
\usepackage[aboveskip=1pt,labelfont=bf,labelsep=period,justification=raggedright,singlelinecheck=off]{caption}

\makeatletter
\renewcommand{\@biblabel}[1]{\quad#1.}
\makeatother

\usepackage{lastpage,fancyhdr,graphicx}
\usepackage{url}
\usepackage{epstopdf}
\usepackage[colorinlistoftodos,prependcaption,textsize=tiny]{todonotes}
\usepackage{regexpatch}
\makeatletter
\xpatchcmd{\@todo}{\setkeys{todonotes}{#1}}{\setkeys{todonotes}{inline,#1}}{}{}
\fancyheadoffset[L]{0in}
\fancyfootoffset[L]{0in}
\graphicspath{ {./images/} }

\usepackage{tikz}
\usepackage{pgfplots}
\pgfplotsset{compat=1.14}

\usepgfplotslibrary{groupplots}

\usepackage{csvsimple}

\definecolor{airforceblue}{rgb}{0.36, 0.54, 0.66}
\definecolor{amaranth}{rgb}{0.9, 0.17, 0.31}
\definecolor{amber}{rgb}{1.0, 0.75, 0.0}
\definecolor{asparagus}{rgb}{0.53, 0.66, 0.42}
\definecolor{burntsienna}{rgb}{0.91, 0.45, 0.32}
\definecolor{antiquefuchsia}{rgb}{0.57, 0.36, 0.51}
\definecolor{beaver}{rgb}{0.62, 0.51, 0.44}

\definecolor{vir1}{rgb}{0.9932    0.9062    0.1439}
\definecolor{vir2}{rgb}{0.2670    0.0049    0.3294}
\definecolor{vir3}{rgb}{0.2812    0.1575    0.4704}
\definecolor{vir4}{rgb}{0.2437    0.2906    0.5381}
\definecolor{vir5}{rgb}{0.1906    0.4071    0.5561}
\definecolor{vir6}{rgb}{0.1471    0.5130    0.5570}
\definecolor{vir7}{rgb}{0.1196    0.6173    0.5368}
\definecolor{vir8}{rgb}{0.2080    0.7187    0.4729}
\definecolor{vir9}{rgb}{0.4249    0.8067    0.3501}
\definecolor{vir10}{rgb}{0.7064    0.8682    0.1715}

\begin{document}
\vspace*{0.2in}

\begin{flushleft}
{\Large
\textbf\newline{Synergistic information in a dynamical model implemented on the human structural connectome reveals spatially distinct associations with age} 
}
\newline
\\
Davide Nuzzi\textsuperscript{1},
Mario Pellicoro\textsuperscript{1},
Leonardo Angelini\textsuperscript{1},
Daniele Marinazzo\textsuperscript{2},
Sebastiano Stramaglia\textsuperscript{1,3}.\\
\bigskip
\textbf{1} Dipartimento Interateneo di Fisica, Università degli Studi Aldo Moro,
Bari and INFN, Sezione di Bari, via Orabona 4, 70126 Bari, Italy\\
\textbf{2} Department of Data Analysis,
Ghent University, 2 Henri Dunantlaan,
9000 Ghent, Belgium\\
\textbf{3} Center of Innovative Technologies for Signal Detection and Processing (TIRES), Università degli Studi Aldo Moro, Bari, via Orabona 4, 70126 Bari, Italy

\bigskip

%
%






\end{flushleft}
\section*{Abstract}
We implement the dynamical Ising model on the large scale architecture of white matter connections of healthy subjects in the age range 4-85 years, and analyze the dynamics in terms of the synergy, a quantity measuring the extent to which the joint state of pairs of variables is projected onto the dynamics of a target one.
We find that the amount of synergy in explaining the dynamics of the hubs of the structural connectivity (in terms of degree strength) peaks before the critical temperature, and can thus be considered as a precursor of a critical transition. Conversely the greatest amount of synergy goes into explaining the dynamics of more central nodes.
We also find that the aging of the structural connectivity is associated to significant changes in the simulated dynamics: 
there are brain regions whose synergy decreases with age, in particular the frontal pole, the Subcallosal area and the Supplementary Motor area; these areas could then be more likely to show a decline in terms of the capability to perform higher order computation (if structural connectivity was the sole variable). On the other hand, several regions in the temporal cortex show a positive correlation with age in the first 30 years of life, i.e. during brain maturation. 


\section*{Introduction}
Recent advances in diffusion imaging and tractography methods allow noninvasive in vivo mapping of white matter cortico-cortical projections at relatively high spatial resolution, thus providing a connection matrix of interregional structural connectivity (SC) representing the {\it geometry} of the brain \citep{sporns2010networks}. 
Dynamical models implemented on the large scale architecture of the human brain may shed light on how function is constrained by the underlying structure. This is the case of the so-called neural mass or mean-field models that describe the collective activity of cell populations \citep{deco2012ongoing}, or phase coupling models \citep{finger2016modeling}, down to abstract models such as the Ising model \citep{deco2012anatomy,haimovici2013brain,marinazzo2014information,stramaglia2017ising}. In particular some studies showed that the resting activity exhibits peculiar scaling properties, resembling the dynamics near the critical point of a second order phase transition \citep{chialvo2010emergent}. Moreover, the possible origin and role of criticality in living adaptive and evolutionary systems has recently been ascribed to adaptive and evolutionary functional advantages \citep{moretti2013griffiths}. 

In this paper we implement a dynamical model on the individual large scale structural connectivity of healthy subjects with age in the range 4-85 years, and look for the dynamical properties of the simulated dynamics which are associated to aging. 

Human aging is the set of characteristics that change over time, signifying someone as older or younger. These changes occur at different hierarchical levels, called metrics of aging: biological aging,  phenotypic aging, and functional aging \citep{ferrucci2018time}. Connectomics \citep{bassett2017network} provides a unique resource for examining how brain organization and connectivity changes across typical aging, both in terms of plasticity and function, and how these differences relate to brain disorders. 
Several studies have used diffusion imaging to investigate changes in structural connectivity over the healthy human lifespan. In general, these studies have observed a non-linear inverted U-shaped trajectory association between age and fractional anisotropy (FA), and an U-shaped trajectory (opposite to FA) for axial, mean, and radial diffusivity \citep{westlye2010life}.
Considering the structural connectivity network provided by diffusion imaging as a complex network, some global metrics of the networks have been found to be correlated with age \citep{wozniak2006advances}. Age showed significant positive correlation to the integrated cost but significant negative correlation to the integrated local efficiency, while showed no significant correlation to the integrated global efficiency \citep{sun2012reorganization}. The reorganization with age of the whole-brain structural and functional connectivity has been described in \citep{betzel2014changes}. The age-related alterations in the topological architecture of the white matter structural connectome has been also studied in \citep{zhao2015age}, where it has been found that  hub integration decreased linearly with age, with loss of frontal hubs and their connections, and that age-related changes in structural connections were predominantly located within and between the prefrontal and temporal modules.  Concerning brain maturation, positive correlation between structural and functional connectivity has been described in \citep{hagmann2010white}, where in addition it has been observed that this relationship strengthened with age. The presence of functional-structural markers which are significantly correlated with aging has been also described in \citep{bonifazi2018structure}. 

The dynamical properties of the simulated activity that we will consider here are derived from the  formalism of information decomposition of target effects from multi-source interactions, i.e. the definition of redundant and synergistic components of the information that a set of source variables provides about a target \citep{lizier2012local}. Applying this framework to the two-dimensional Ising model as a paradigm of a critically transitioning system, and disentangling the components of the information both at the static level and at the dynamical one, 
it has been recently shown that a key signature of an impending phase transition (approached from the disordered side) is the evidence that the synergy peaks in the disordered phase, both considering only instantaneous interactions, and also lagged ones: the synergy can thus be considered a precursor of the transition \citep{marinazzo2019synergy}.
Here we implement the Ising model on the real structural geometry of brain here estimated from Diffusion Tensor Imaging (DTI). On one hand we find, as the temperature is lowered, that the synergy can still be considered as a precursor of transition in an inhomogeneous network and not only in an ordered lattice: the synergy towards the most connected brain regions (averaged over pairs of drivers) peaks before the maximum of the susceptibility, which is conventionally taken as the transition point in finite size systems. On the other hand, we find that the synergy towards some regions in the temporal cortex is positively correlated with age, whilst the synergy in the frontal pole, and in the subcallosal area, is negatively correlated with age. 
Intuitively, synergy is the information about the target variable that is uniquely obtained taking the sources together, but not considering them alone; hence it measures to what extent the activity of given target region of the brain is the projection of the joint activity of two other driver regions, in other words the capability to perform computation of higher level. In this sense, these results suggest that the function of some regions of the brain may deteriorate with age with a contribution from the changing white matter geometry, whilst other regions receive more synergistic information as age increases.

\section*{Materials and methods}

	\subsection*{Data}
	In this work we analyze data form the NKI/Rockland life-span study~\citep{nooner2012nki}, in particular the already processed connectome data provided by the USC Multimodal Connectivity Database~\citep{brown2012ucla},
consisting of 196 connectomes of healthy subjects based on 3T dMRI acquisition (voxel size, 2 mm\textsuperscript{3}; $64$ 
	gradient directions; TR, 10000 ms; TE, 91 ms; further details are provided in~\citep{brown2012ucla}). 
	The resulting structural connectivity matrices $J_{ij}$ consist of $N = 188$ regions of interest (ROI), obtained using the Craddock atlas~\citep{craddock2012whole}, 
	linked together by weighted connections based on the number of streamlines connecting pairs of ROIs. The matrix $J_{ij}$ is symmetric by 
	construction  thus giving rise to a weighted undirected graph.
	The age of subjects
	ranges from $4$ to $85$ years.
	We  consider the individual structural connectivity matrices, and the average connectivity matrix $J_{ij}^\text{avg}$ obtained averaging over all subjects.
	
	
\subsection*{Theoretical framework and implementation}


Firstly we implement the Ising model, with Glauber dynamics, on the average connectome network $J_{ij}^\text{avg}$, using the following Hamiltonian:
\begin{eqnarray}
\label{eq:Hamiltonian}
    \mathcal{H} = -\frac{1}{2} \sum_{i,j = 1}^N J_{ij} s_i s_j,
\end{eqnarray}
and updating rules given by 
\begin{equation}
    p(s_i \rightarrow -s_i) = \frac{1}{1 + \exp\left(\beta \Delta E_i\right)},
\end{equation}
where $\Delta E_i = 2 s_i \sum_j J_{ij} s_j$ represents the variation of the total energy of the system following the flip of the spin $s_i$. Varying the inverse temperature $\beta$, the  susceptibility shows a peak that identifies  the {\it critical state} of this finite size system, related to a phase transition occurring in the limit of large networks~\citep{dorogovtsev2008critical}. It is worth mentioning that recently this stretching of criticality (from a single point to a more relaxed regime) observed in dynamical models defined on brain networks, has been described in the frame of Griffith's phases \citep{moretti2013griffiths}.
Simulations are initially run for a relaxation time of  $10^5$ updates for the first value of $\beta$, then we start the following procedure: we vary the temperature adiabatically, discarded the first  $10^4$ updates and then collected the following  $10^6$ updates for statistics. The procedure is repeated for $20$ runs at each of the 80 temperature points enclosing the phase transition. We make sure that the state of each spin is updated exactly one time for each iteration of the Glauber dynamics; this allows us to collect one time series $s_i(t)$ for every spin, where $t$ is a discrete time index running from $0$ to the total simulation time. In order to characterize the flow of information between the time series in the context of information theory, we interpret each $s_i(t)$ as a single realization of a discrete-time stationary stochastic process $S_i(t)$, but for the sake of simplicity we will use the notation $s_i(t)$ also to refer to the stochastic process.
Joint probabilities used in the calculations of information quantities are obtained  directly from the data samples as the frequency of each configuration. \\
The flow of information between variables can be measured in the framework of information dynamics using the transfer entropy (TE), a quantity introduced in~\citep{schreiber2000measuring} and based on appropriate conditioning of transition probabilities~\citep{bossomaier2016introduction}. Unlike mutual information $I(X;Y)$, a symmetric quantity which measures only  the information that is statistically shared by the variables $X$ and $Y$, transfer entropy is an inherently asymmetrical quantity and can effectively distinguish between driving and responding elements. Let $s_i$ be the stochastic process associated to a given spin, taken as the target variable, and let $\tilde{s}_i$ be the same process one time step in the future. Let $\{s_j, s_k, \dots, s_r\}$ be a group of spins that are assumed to be drivers for the spin $s_i$. The transfer entropy from the group of drivers to the target is defined as
\begin{eqnarray}
\label{eq:TE}
    \mathcal{T}_{\{jk\dots r\} \rightarrow i} = I \big( \tilde{s}_i; \{s_j, \dots, s_r\} \big| s_i \big).
\end{eqnarray}
The choice of different groups of source variables leads to the definition of various information quantities that are commonly found in the literature. If we choose only one source variable $s_j$ and we average over all possible sources and targets we get the pairwise transfer entropy; it has been shown that this measure peaks at the critical point both for the Ising model on a regular 2D lattice and on brain graphs~\citep{barnett2013information,marinazzo2014information}. If we choose all the spins but $s_i$ as sources and then average over all possible targets we get the global transfer entropy (GTE). Information flow, as quantified by GTE, peaks in the paramagnetic phase and is thus able to predict an imminent transition~\citep{barnett2013information}. Unfortunately numerical estimation of this quantity is unfeasible in systems involving a large number of driver for each target. 
As discussed in~\citep{lizier2010information} GTE is a measure of collective information transfer, capturing both pairwise and higher-order (multivariate) correlations to a variable. It follows that explicitly disentangling the components of the collective information flow is needed to get a better description of the system in the proximity of the transition. As it has been demonstrated in~\citep{marinazzo2019synergy} considering as few as two sources at the same time is sufficient to construct a precursor of the transition. Let $s_i$ be the target spin and $s_j, s_k$ be two different driver spins, then the desirable information decomposition is:
\begin{eqnarray}
\label{eq:InformationDecomposition}
    \mathcal{T}_{jk \rightarrow i } &=& \mathcal{U}_{j \rightarrow i} + \mathcal{U}_{k \rightarrow i} + \mathcal{R}_{jk \rightarrow i} + \mathcal{S}_{jk \rightarrow i} ,\\
    \mathcal{T}_{j \rightarrow i } &=& \mathcal{U}_{j \rightarrow i} + \mathcal{R}_{jk \rightarrow i} , \\
    \mathcal{T}_{k \rightarrow i } &=& \mathcal{U}_{k \rightarrow i} + \mathcal{R}_{jk \rightarrow i} .
\end{eqnarray}
The terms $\mathcal{U}_{j \rightarrow i}$ and $\mathcal{U}_{k \rightarrow i}$ quantify the components of the information about the future of spin $s_i$ that are unique to the sources $s_j$ and $s_k$ respectively, i.e. the information that can be acquired considering each source as the only driver. The term $\mathcal{R}_{jk \rightarrow i}$ describes the redundant information, the component that is shared between the two source variables. Finally, the term $\mathcal{S}_{jk \rightarrow i}$ quantifies the synergy between the sources, intended as the amount of information that can only be acquired considering the sources together, but not considering them alone. Synergy is the only term in this decomposition that contains higher-order correlations that can't be captured by pairwise measures.

Shannon theory of information does not include a definition for synergy and redundancy, in fact there are many different conceptual definitions for those two quantities~\citep{harder2013bivariate,quax2017quantifying,griffith2014intersection}. The information decomposition provided in Eq. \ref{eq:InformationDecomposition} contains only three equations for the four unknown quantities $\mathcal{U}_{j \rightarrow i},\mathcal{U}_{k \rightarrow i},\mathcal{R}_{jk \rightarrow i},\mathcal{S}_{jk \rightarrow i}$, therefore the definition of only one of them is needed in order to solve the system. The so-called minimum mutual information (MMI) PID~\citep{barrett2015exploration} gives a definition for the redundancy: it assumes that the information that is shared between the sources, independently of the correlations between them, can be identified with the minimum of the information provided by each individual source to the target. Specifically:
\begin{equation}
    \mathcal{R}_{jk \rightarrow i} = \min \left\{\mathcal{T}_{j \rightarrow i}, \mathcal{T}_{k \rightarrow i} \right\}    
\end{equation}
Another choice for the information decomposition is given in~\citep{bertschinger2014quantifying} and is based on the following idea. Let $p(s_i, s_j, s_k)$ be the joint probability distribution of the three spins and let $p(s_i,s_j),p(s_i,s_k)$ be the two marginal distributions that involve the target and only one driver. Then the redundancy and the unique information should only depend on the marginal distributions but not on the particular choice of $p(s_i, s_j, s_k)$, that should only affect the synergy. If we define $\Delta_p$ as the set of all the trivariate probability distributions $q$ which give rise to the same marginal distributions as $p$, i.e. $q(s_i, s_j) = p(s_i,s_j)$, $q(s_i,s_k) = p(s_i,s_k)$, we can define the synergy as
\begin{equation}
    \mathcal{S}_{jk \rightarrow i} = \mathcal{T}_{jk \rightarrow i} - \min_{q \in \Delta_p}  \tilde{\mathcal{T}}_{jk \rightarrow i},
\end{equation}
where $\tilde{\mathcal{T}}$ is the transfer entropy evaluated using $q$ as the joint probability distribution.
It has been shown~\citep{barrett2015exploration} that in the case of Gaussian stochastic variables those two approaches are equivalent and provide the same decomposition. We have verified numerically that the same holds true for the problem at hand.\\

The average connectome is a dense network, hence for the dynamics on the average connectome  we take into account all the triplets of brain nodes and evaluate the information decomposition for each triplet; in order to evaluate  the average incoming synergy for each node, $\mathcal{S}$ is averaged over all the triplets having that node as a target.

The individual structural networks, on the other hand, have many nearly zero entries; therefore we make a selection of relevant triplets of brain nodes $\{j,k,i\}$, with target i, requiring that both $J_{ji}$ and $J_{ki}$ are higher than a threshold $J_{th}$. We fix $J_{th}$ so as the total number of considered links is 20$\%$ of all the possible pairs, and verify that our results are robust w.r.t. the choice of $J_{th}$. For all the selected triplets we evaluate the information decomposition and, in order to evaluate the typical incoming synergy for each node of an individual network, we average the synergy over all the  triplets, among the selected ones, which have that node as the target. 
\begin{figure}[!h]
    \centering
    \includegraphics{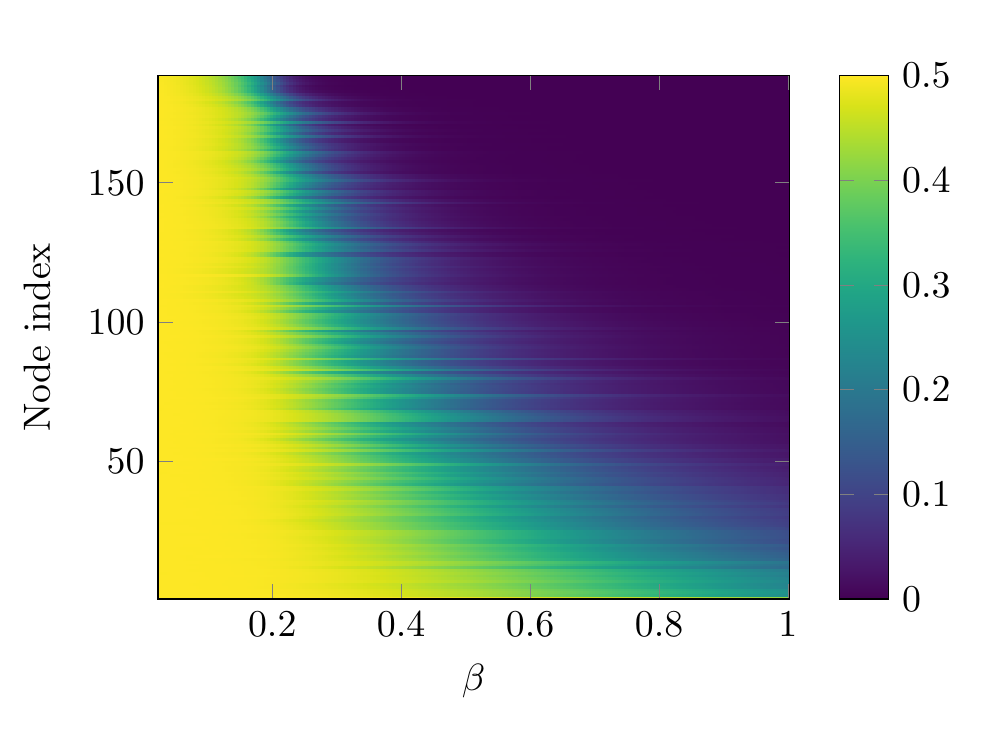}
    \caption{The probability of flip, for each spin at one time step, is depicted as a function of $\beta$. The nodes are ordered according to increasing strength of the average connectome. At low temperatures spins {\it freeze}, each spin freezing at a different temperature; the figure shows that the strength of nodes explains, to a certain extent, the variability of the freezing temperature.}
    \label{fig:FlipProb}
\end{figure}

\section*{Results}
{\bf Ising model on the average connectivity network.}

Because of the range of strengths of each node, each spin undergoes a dynamical transition at a different temperature. This phenomenon can be visualized studying the probability for the spin to flip at each time step, shown in Figure~\ref{fig:FlipProb}. For $\beta = 0$ every spin has exactly a 50\% chance to flip, whilst as $\beta$ gets larger the probability drops to zero at a different rate, depending on the strength of the corresponding node; ordering the nodes by their strength shows that stronger nodes tend to freeze at higher temperatures. As already noted in \citep{marinazzo2014information}, as the temperature is lowered the sub-network constituted by the hubs (we verified the average connectivity network is a rich club) tend to align and to build a cluster of highly correlated spins with slow dynamics. More and more spins are recruited to join this cluster as the temperature is further lowered, until all the system is magnetized. As a consequence, this system does not show a critical temperature, rather it is characterized by a critical range of temperatures, in accordance with Griffith's theory \citep{moretti2013griffiths}.


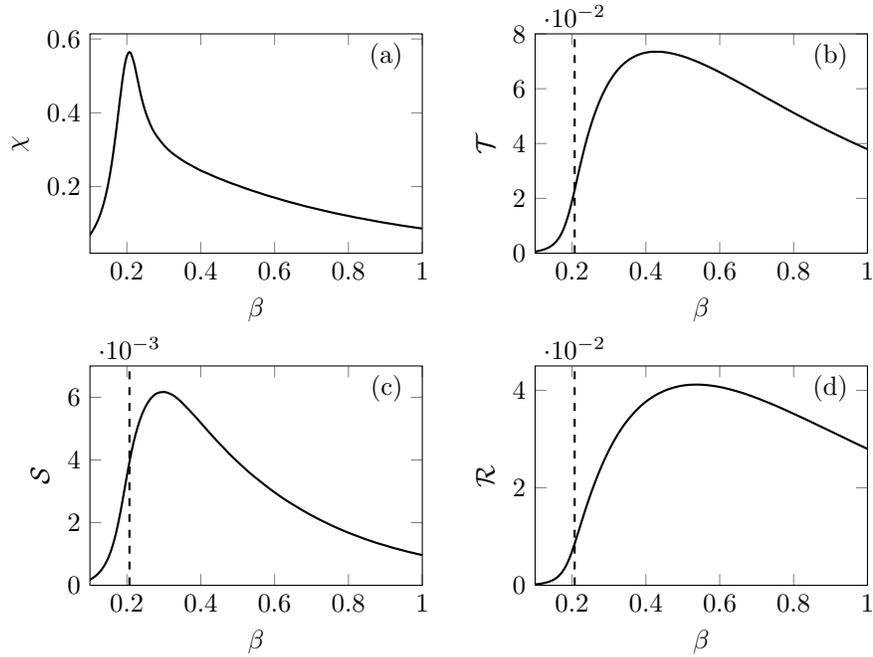
\begin{figure}[!h]
\centering
\begin{tikzpicture}

\begin{groupplot}
[   
    group style={group name=my plots,group size= 2 by 2, 
    vertical sep=1.5cm, horizontal sep = 1.5cm },height=4.5cm, width=6cm,
    xmin = 0.1, xmax = 1, xlabel = $\beta$, ylabel near ticks
]

\nextgroupplot[ylabel = $\chi$]

    \addplot [thick] table [x=beta, y=chi, col sep=comma, mark = none] {./data/MeanData.dat};
\node [below left] at (rel axis cs:0.97,1) {(a)};

\nextgroupplot[ymin = 0 , ymax = 0.08, ylabel = $\mathcal{T}$]

    \addplot [thick] table [x=beta, y=te_mean, col sep=comma, mark = none] {./data/MeanData.dat};
	\addplot[thick, samples=50,domain=0.1:0.4, black, dashed] coordinates {(0.207,0)(0.207,0.08)};
\node [below left] at (rel axis cs:0.97,1) {(b)};

\nextgroupplot[ymin = 0 , ymax = 0.007, ylabel = $\mathcal{S}$]

    \addplot [thick] table [x=beta, y=synergy_mean, col sep=comma, mark = none] {./data/MeanData.dat};
	\addplot[thick, samples=50,domain=0.1:0.4, black, dashed] coordinates {(0.207,0)(0.207,0.007)};
\node [below left] at (rel axis cs:0.97,1) {(c)};

\nextgroupplot[ymin = 0 , ymax = 0.045, ylabel = $\mathcal{R}$]

    \addplot [thick] table [x=beta, y=redundancy_mean, col sep=comma, mark = none] {./data/MeanData.dat};
	\addplot[thick, samples=50,domain=0.1:0.4, black, dashed] coordinates {(0.207,0)(0.207,0.045)};
\node [below left] at (rel axis cs:0.97,1) {(d)};

\end{groupplot}

\end{tikzpicture}
\caption{{\bf Information decomposition for the Ising model on the brain connectome.} 
(a) Susceptibility $\chi$ versus the inverse temperature $\beta$; the temperature where it peaks is considered as the critical state of the system. (b)(c)(d) Trivariate transfer entropy, redundancy and synergy averaged over all spin triplets, depicted against $\beta$, peaking in the ferromagnetic state. The dashed line indicates the critical temperature.}
\label{fig:Average}
\end{figure}

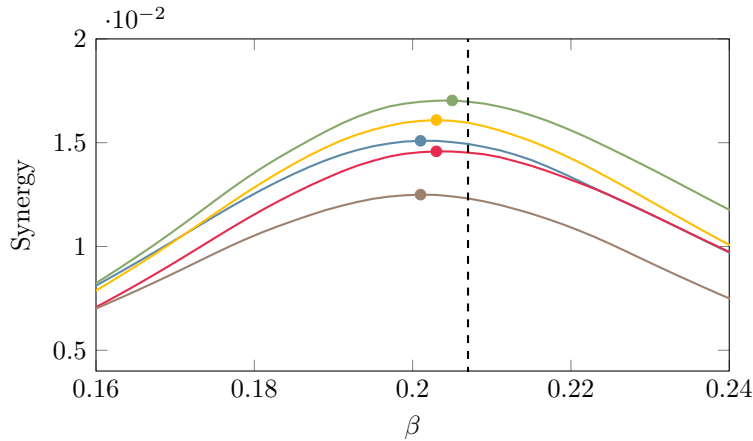
\begin{figure}[!h]
\centering
\begin{tikzpicture}
\begin{axis}[xmin = 0.16, xmax = 0.24, ymin = 0.004, ymax = 0.02,
             xlabel = $\beta$, ylabel = Synergy, 
             width = 10cm, height = 6cm, 
             xtick distance=0.02]
             
    \addplot [thick, airforceblue] table [x=beta_s, y=synergy_smooth_1, col sep=comma, mark = none] {./data/Synergy_first.dat};
    \addplot [thick, amaranth] table [x=beta_s, y=synergy_smooth_2, col sep=comma, mark = none] {./data/Synergy_first.dat};
    \addplot [thick, amber] table [x=beta_s, y=synergy_smooth_3, col sep=comma, mark = none] {./data/Synergy_first.dat};
    \addplot [thick, asparagus] table [x=beta_s, y=synergy_smooth_5, col sep=comma, mark = none] {./data/Synergy_first.dat};
    \addplot [thick, beaver] table [x=beta_s, y=synergy_smooth_6, col sep=comma, mark = none] {./data/Synergy_first.dat};  
    
	\addplot[thick, samples=50,domain=0.1:0.4, black, dashed] coordinates {(0.207,0)(0.207,0.02)};
    	
    \addplot[mark=*, airforceblue] coordinates {(0.201,0.0150908535679259)};
    \addplot[mark=*, amaranth] coordinates {(0.203,0.014582389420118)};
    \addplot[mark=*, amber] coordinates {(0.203,0.0160927494962468)};
    \addplot[mark=*, asparagus] coordinates {(0.205,0.0170363801369974)};
    \addplot[mark=*, beaver] coordinates {(0.201,0.0124971546394868)};
	
\end{axis}
\end{tikzpicture}
\caption{The synergy towards the five brain nodes with highest strength is depicted versus $\beta$; their plot peaks before the critical temperature here identified as the peak of susceptibility.  }
\label{fig:SynergyFirst}
\end{figure}

In figure \ref{fig:Average} we have depicted, as a function of the coupling $\beta$, the global quantities related to the information decomposition, i.e. the trivariate transfer entropy, the average synergy and the redundancy, as well as the susceptibility $\chi$, the classical index of criticality. Looking at the figure, we see that the peak of  $\chi$ precedes, as $\beta$ increases, the peaks of the synergy, the transfer entropy and the redundancy in order. The peak of $\chi$ in this heterogeneous system occurs at lower $\beta$ than the peak of the synergy, differently from what happens in the Ising model on a regular lattice \citep{marinazzo2019synergy}; however one may consider the synergy for each target node, averaged over the pairs of driving variables, and realize that each node experiences its maximum of the synergy at different temperatures. As depicted in figure \ref{fig:SynergyFirst}, the hubs  show their peak of synergy before the peak of $\chi$, hence their synergy can be seen as precursor of the transition to order.

\begin{figure}[!h]
\centering
\includegraphics[width = 0.7\linewidth]{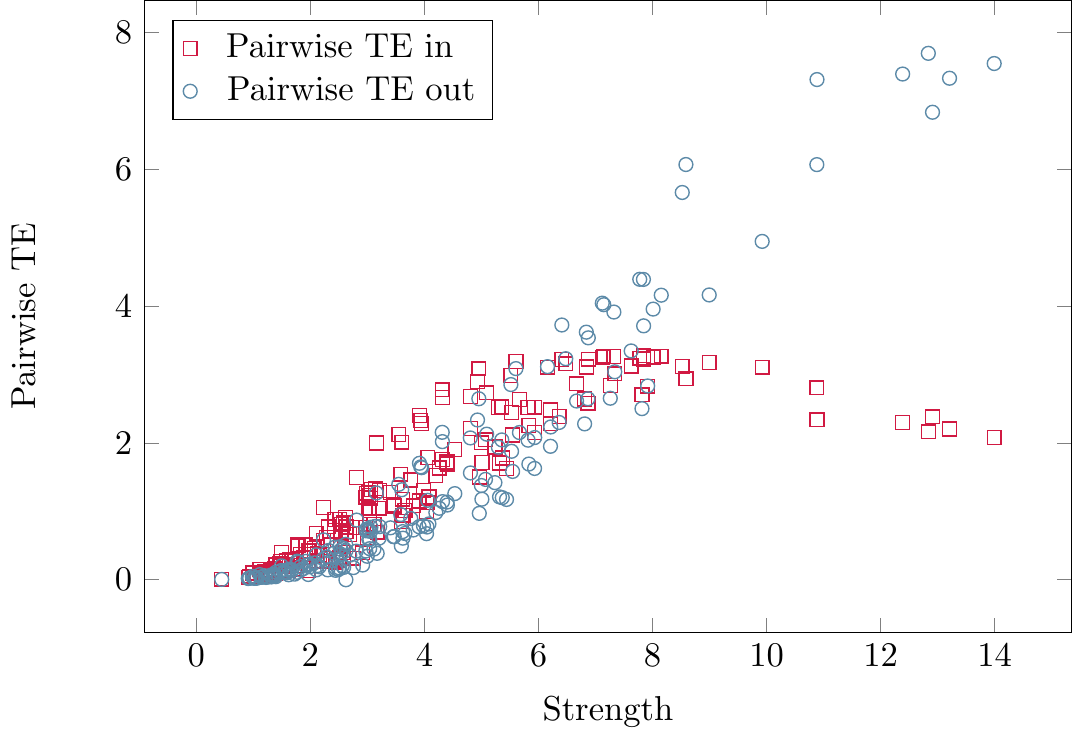}
\caption{{\bf Incoming and outgoing transfer entropy.}
The amount of information that a fixed driver node transfers to all possible targets (outgoing TE) increases steadily with the strength of the node. On the other hand, the amount of information received by a target node from all the drivers (incoming TE), behaves similarly for small strength values but decreases above a certain strength threshold.
}
\label{fig:TEpair}
\end{figure}

Another interesting remark to be made is that, as noticed e.g. in \citep{novelli2019deriving}, the pairwise transfer entropy from a source to a target node in a network does not depend solely on the local source-target link weight, but on the wider network structure that the link is embedded in. Deeply connected is the fact that the information flow in networks obey the law of diminishing marginal returns, see \citep{marinazzo2012information}: in figure \ref{fig:TEpair} we depict the incoming and outgoing transfer entropy of each node, as a function of the strength of the node, where the temperature is fixed at the peak of  $\chi$. Due to the limited capacity of a spin to encode the incoming information, nodes with high strength send more information to the system than it gets; nodes with low strength send and receive approximately the same amount of information. These findings support the picture of a dynamics shaped by the hubs of structural connectivity. 

\begin{figure}[!h]
\centering
\begin{tikzpicture}

\begin{groupplot}
[   
    group style={group name=my plots,group size= 3 by 1, 
    vertical sep=1.5cm, horizontal sep = 0.1cm },height=5cm, width=6cm,
     ylabel near ticks
]

\nextgroupplot[ylabel = Incoming synergy, xlabel = Strength]

    \addplot [only marks, mark = o, airforceblue, mark size = 2]  table [x=strength, y=syn_in, col sep=comma] {./data/SynergyCorr.csv};

\nextgroupplot[xlabel = Betweenness, ymajorticks=false]

    \addplot  [only marks, mark = o, airforceblue, mark size = 2]    table [x=betweenness, y=syn_in, col sep=comma] {./data/SynergyCorr.csv};

\nextgroupplot[xlabel = Closeness, ymajorticks=false,
xticklabel style={
        /pgf/number format/fixed,
        /pgf/number format/precision=3
},
scaled x ticks=false]

    \addplot  [only marks, mark = o, airforceblue, mark size = 2]   table [x=closeness, y=syn_in, col sep=comma] {./data/SynergyCorr.csv};

\end{groupplot}

\end{tikzpicture}
\vspace{0.5cm}
\caption{{\bf Comparing synergy with topological indices.} 
From the left to the right, for each brain node the incoming synergy at criticality is compared with the strength of nodes, the betweenness and the closeness.}
\label{fig:SynergyCorrelations}
\end{figure}
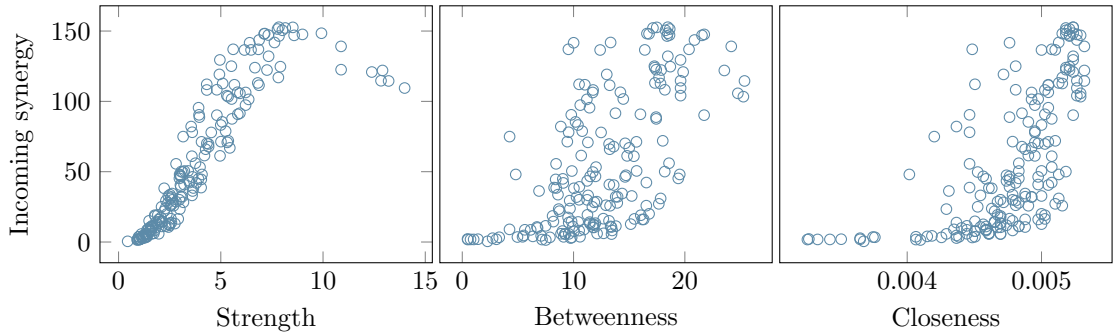

Here we show that a similar behaviour holds for the average incoming synergy to each given node, in other words when we average the synergy of all the triplets with the given node as the target. In figure (\ref{fig:SynergyCorrelations}) the incoming synergy, at criticality, is depicted versus the topological character of nodes: strength, betweeness and closeness. We see that the hubs of structural connectivity, which as described above are the drivers of the dynamics, are not among the nodes towards which synergy is highest. Measures of centrality, like betweeness and closeness, appear to be more associated to synergy than strength: the joint state of brain regions is more likely to be projected in {\it central} regions of the brain.

\begin{figure}[!h]
\centering
\includegraphics[width = 0.6\linewidth]{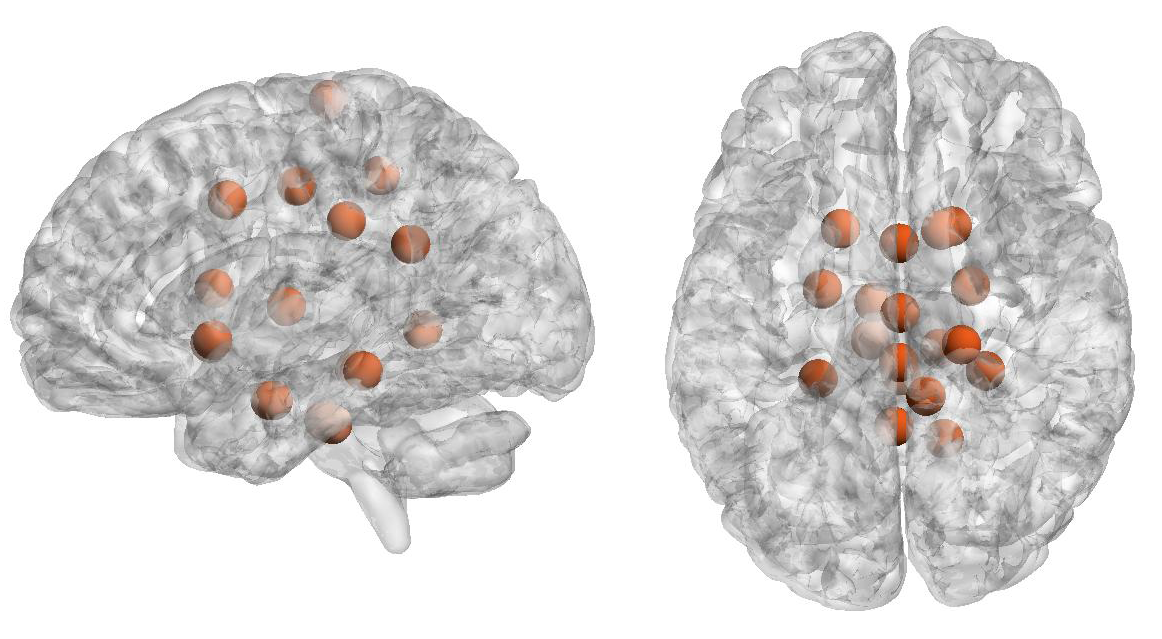}
\caption{{\bf Hubs of synergy.}
Top nodes for the value of incoming synergy, radius and color of the spheres are arbitrary:\\
'Right Hippocampus',
'Brain Stem',
'Right Parahippocampal posterior'
'Left Parahippocampal posterior',
'Right Cingulate posterior',
'Right Precentral',
'Left Thalamus',
'Left Parahippocampal posterior',
'Left Hippocampus',
'Right Lingual',
'Right Caudate',
'Right Cingulate anterior'
}
\label{fig6}
\end{figure}

In figure (\ref{fig6}) we have depicted the 10$\%$ nodes with highest average incoming synergy. Notably these nodes correspond to Hippocampus and Parahippocampal regions, which are supposed to play important role in higher order cognitive functions, specifically learning and memory processes \citep{nemanic2004hippocampal}; the Brainstem, which controls the flow of messages between the brain and the rest of the body, as well as several body functions; the Cingulate Cortex, which has been associated to several complex cognitive functions, such as empathy, impulse control, emotion, and decision-making; the Thalamus, whose main function is to relay motor and sensory signals to the cerebral cortex.

These two figures were plotted with BrainNet Viewer \citep{xia2013brainnet}.

{\bf Ising model on individual connectivity networks and correlations with aging.}

\begin{figure}[!h]
\centering
\includegraphics[width = 0.8\linewidth]{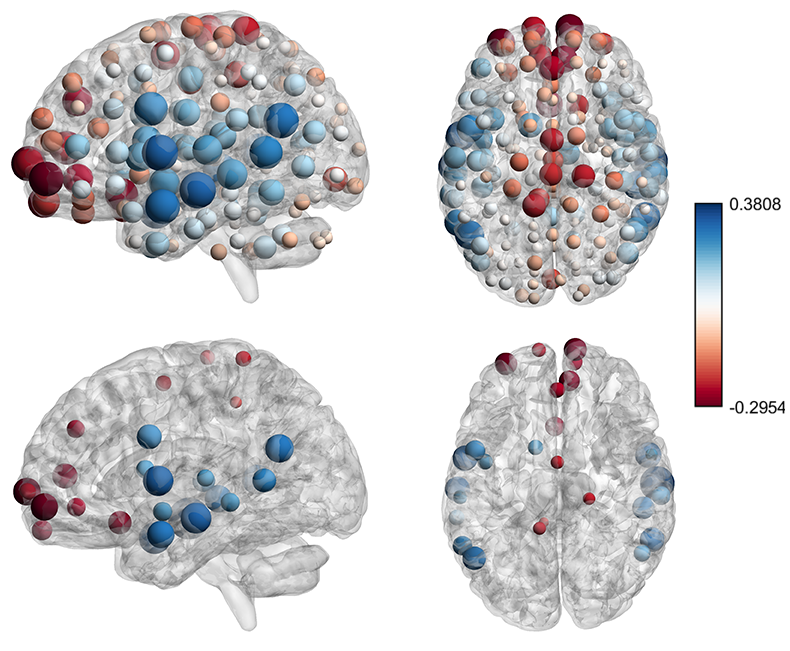}
\caption{{\bf Correlations between synergy and age.}
(Top) Spearman correlation coefficient between the chronological age of the patient and the incoming synergy for each node, i.e. the average synergy for all triplets having that node as a target. Both positive (blue) and negative (red) correlations are found. The radius of the spheres is proportional to the absolute value of the correlation.
(Bottom)  Only the regions with significant correlation are shown, evaluated using Hochberg adjusted p values at a significance level $\alpha = 0.05$. Synergy is positively correlated with age in the following regions:
'Left Middle Temporal posterior',
'Right Middle Temporal anterior',
'Left Angular',
'Right Superior Temporal posterior',
'Left Middle Temporal temporooccipital',
'Right Middle Temporal posterior',
'Left Middle Temporal temporooccipital',
'Left Middle Temporal anterior',
'Left Central Opercular',
'Right Superior Temporal posterior'
;
and negatively correlated in 
'Left Frontal Pole',
'Left Subcallosal   ',
'Right Frontal Pole', 'Paracingulate sulcus', 'Juxtapositional Lobule'
.}
\label{age}
\end{figure}

The implementation of the Ising model on the individual networks of healthy subjects shows the robustness of the scenario described above, indeed qualitatively similar results have been found on individual networks, each network showing, however, its own critical temperature as evaluated at the peak of the susceptibility. Remarkably, this analysis provides an individual pattern of synergy, evaluated at criticality, so as to evidence those nodes whose synergy co-varies significantly with age.  In figure (\ref{age}) we depict the Spearman correlation coefficient between incoming entropy and age, for each brain node, showing a symmetric pattern. 
We choose measures of association which are robust to the presence of outliers, and suitable for multiple comparisons. We use the implementations contained in the robust statistical toolbox by Rand Wilcox (Rallfun-v37.txt, update September 2019), using R version 3.5.3 (R Core Team, 2018). In particular we adopt the skipped Spearman correlation with adjusted p-values in conjunction with Hochberg's method to control family-wise error. This approach is referred to as L3 in \citep{wilcox2018improved}, and implemented in the R function scorregciH.
This analysis yields twenty-nine regions whose synergy is significantly associated with aging at a corrected $\alpha$ value of 0.05, seventeen regions positively correlated with age and twelve negatively correlated with age, depicted in figure (\ref{age}).

\begin{figure}[!h]
\centering
\includegraphics[width = \linewidth]{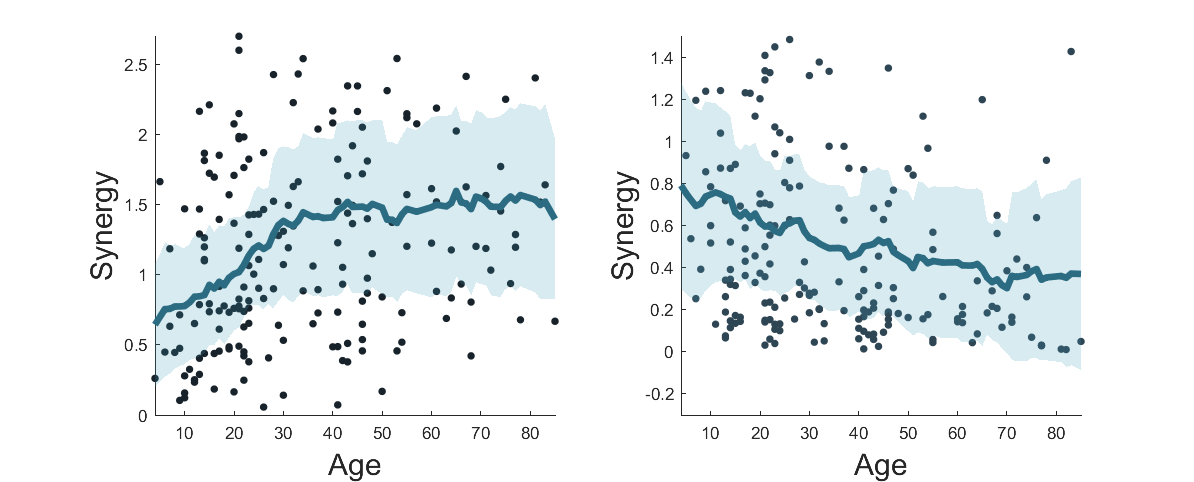}
\caption{{\bf Scatter plot of synergy and age for two representative brain regions.}\\
Left: Right Superior Temporal posterior, positive correlation.\\
Right: Right Frontal pole, negative correlation.\\
Local average and standard deviation are evaluated using the first $20$ neighbours of each point.
}
\label{aging}
\end{figure}

The nodes whose synergy is significantly decreasing with age are frontal poles and subcallosal regions, as well as Paracingulate Sulcus and Juxtapositional Lobule (formerly Supplementary motor area). It has been suggested that the frontal lobes are the part of the brain most profoundly affected by the aging process \citep{tisserand2002regional}; a review of age related changes in MR Spectroscopy, Functional MRI, and Diffusion-Tensor Imaging can be found in \citep{minati2007mr}.  Successful cognitive aging and its functional imaging correlates are discussed in \citep{eyler2011review}.   Juxtapositional Lobule is among the cortical motor regions whose dynamic has been shown to be modulated by age in \citep{wang2019aging}. The nodes whose synergy is significantly increasing with age are mostly located in the temporal cortex. In figure (\ref{aging}) we have depicted for example the synergy versus age for two brain nodes, the right superior temporal posterior (showing a significant positive correlation with age) and the right frontal pole (showing a negative significant correlation with age). Figure (\ref{aging}) suggests that some regions show a slow and continuous decrease in synergy. On the other hand the brain regions with a significant positive correlation actually show an increase in the synergy in the first decades of life, and a plateau later: hence such an increase of synergy could be associated to brain maturation rather than to a compensatory effect due to aging.



%

\section*{Conclusions}
We have implemented the Ising model on the large scale structural geometry of the brain as measured by DTI, for individual networks of healthy subjects with different age, as well as on the average connectome, in order to analyze the simulated neural dynamics in terms of the synergy, a quantity which has recently been introduced in the field of partial information decomposition of information and that, in the case of the brain, measures to what extent the joint state of pairs of brain regions is projected onto the dynamics of a target region.

Compared to the behaviour of synergy of the Ising model on the regular two dimensional lattice \citep{marinazzo2019synergy}, in the present case this quantity still can be considered as a precursor of the transition but in a local sense w.r.t. the hubs of the strength. Indeed the transition to order here is orchestrated by the hubs of the structural connectivity, and we have shown that the synergy of these hubs peaks  at higher temperature than the critical temperature, therefore the main findings of \citep{marinazzo2019synergy} can be generalized to heterogeneous networks provided that the synergy of structural hubs is considered.
It is remarkable that the hubs of the synergy are not hubs of the strength, rather they are central nodes of the network, although neither the betweness nor the closeness are fully monotonical with the synergy.

Implementing the Ising model on the individual structural connectomes, we found that there are brain regions whose incoming synergy decreases continuously with age, in particular the frontal pole and the subcallosal area: these brain nodes are known to be affected by the aging process \citep{kievit2014distinct}. 
Other brain regions, in the temporal cortex, show a positive correlation with age; however looking at the scatter plot suggests that the increase stabilizes around the age of 30. As a consequence these results suggest that the temporal cortex experience a remarkable maturation in term of the synergy, i.e. on the capability of performing higher order computation. Most results about brain maturation refer to subcortical regions and prefrontal regions: the analysis of synergy here seem to suggest a role of the temporal cortex.

Of course from the neuro-physiological point of view, our results have limited interpretability. Our analysis only takes into account the variation with age of the structural geometry, and employs an abstract model of spins to mimic the neural activity. 
However, following the line of similar studies, we believe that simulation of abstract or realistic dynamical systems, on individual brain geometry, is a promising field which can provide potentially useful insights.

\section*{Data and code availability}
The processed connectomes are retrieved from \url{http://umcd.humanconnectomeproject.org} (NKI-Rockland).
The code to simulate the dynamics of the Ising spins coupled according to the connectomes, as well as the synergy values for each region and each subject, are available at \url{https://github.com/danielemarinazzo/ising_synergy_brain}.


\bibliographystyle{chicago} 
\bibliography{references}

%
%
%

\end{document}